%% file: photoz-dr8.tex
\documentclass[preprint]{aastex}

\usepackage{verbatim}
\usepackage{color}
\usepackage[normalem]{ulem} % for striking out with \sout
\input{newcommands.tex}

\newcommand{\localbias}{10-15\%}

\newcommand{\modelrmin}{15.0}
\newcommand{\modelrmax}{29.0}
\newcommand{\rmin}{15.0}
\newcommand{\rmax}{21.8}
\newcommand{\pofz}{$P(z$)}
\newcommand{\pofzw}{$P(z)_w$\ }
\newcommand{\nofz}{$N(z$)}
\newcommand{\nwei}{N(z)_{\rm wei}}
\newcommand{\npz}{N(z)_{\rm P}}

\newcommand{\contamworst}{10\%}
\newcommand{\nphoto}{58,533,603}

\newcommand{\matchrad}{2\arcsec}

\newcommand{\downloadURL}{{\tt http://www.sdss3.org/dr8/data\_access.php\#VAC}}

\newcommand{\acronym}{ProbWTS}
\def\eps@scaling{1.0}% 

\slugcomment{Last revision \today}
\shortauthors{Sheldon}
\shorttitle{DR8 Photoz Catalog}

\begin{document}

\title{Photometric Redshift Probability Distributions \\for Galaxies in the SDSS DR8}

\input{authors.tex}

\begin{abstract}

We present redshift probability distributions for galaxies in the SDSS DR8
imaging data.  We used the nearest-neighbor weighting algorithm presented in
\citet{LimaPhotoz08} and \citet{CunhaPhotoz09} to derive the ensemble redshift
distribution \nofz, and individual redshift probability distributions \pofz\
for galaxies with \rmag$ < $\rmax.  As part of this technique, we calculated
weights for a set of training galaxies with known redshifts such that their
density distribution in five dimensional color-magnitude space was proportional
to that of the photometry-only sample, producing a nearly fair sample in that
space.  We then estimated the ensemble \nofz\ of the photometric sample by
constructing a weighted histogram of the training set redshifts.  We derived
\pofz s for individual objects using the same technique, but limiting to
training set objects from the local color-magnitude space around each
photometric object.  Using the \pofz\ for each galaxy, rather than an ensemble
\nofz, can reduce the statistical error in measurements that depend on the
redshifts of individual galaxies. The spectroscopic training sample is
substantially larger than that used for the DR7 release, and the newly added
PRIMUS catalog is now the most important training set used in this analysis by
a wide margin.  We expect the primary source of error in the \nofz\
reconstruction is sample variance: the training sets are drawn from relatively
small volumes of space.  Using simulations we estimated the uncertainty in
\nofz\ at a given redshift is $\sim$\localbias.  The uncertainty on
calculations incorporating \nofz\ or \pofz\ depends on how they are used; we
discuss the case of weak lensing measurements.  The \pofz\ catalog is publicly
available from the SDSS website.  

\end{abstract}

\section{Introduction} \label{sec:intro}

Photometric redshifts are estimates of redshift derived using broad-band
photometric observables such as magnitudes and colors \citep{bau62,pus82,koo85,loh86,con95}. 
Typically, the set of observables for a given galaxy are not sufficient to uniquely specify its
redshift, but only a probability distribution, the \pofz.  These \pofz s are
often relatively broad. For simplicity of use and interpretation, one commonly
uses a single number, the photometric redshift, as the best estimate of a
galaxy's redshift.  As several recent works have shown
\citep{man08,CunhaPhotoz09,wit09,bor10,abr11}, the use of a single number to
represent the \photoz\ leads to biases.  Working with the full \pofz\ for each
galaxy yields better estimates of the overall redshift distribution, $N(z)$,
and can decrease biases in cosmological analyses.  We note that several public
\photoz\ codes exist that can produce a \pofz\ per galaxy, e.g.  {\it Le Phare}
\citep{arn99,ilb06}, {\it ZEBRA} \citep{fel06}, {\it BPZ} \citep{coe06}, {\it
ArborZ} \citep{ger10}, and our own method \citep{CunhaPhotoz09}, henceforth
referred to as \acronym, which is an acronym for Probability Distributions from
Weighted Training Sets.  We use \pofzw when referring to the \pofz\  derived from
\acronym.

In this paper, we describe a \pofz\ catalog for objects detected in the Data
Release 8 \citep[SDSS DR8;][]{dr8} of the Sloan Digital Sky Survey III
\citep[SDSS III;][]{Eisenstein2011}.  We use the method of
\citet{CunhaPhotoz09}, which was also applied to SDSS DR7 \citep{dr7}, with
improvements in the training set and photometry.  The DR7 catalog of
\cite{CunhaPhotoz09} has been successfully used in cosmological analyses,
allowing, for example, for the first measurement of the transverse BAO scale
derived purely from angular information, i.e. without using the 3D
power-spectrum \citep{car11}  and for the measurement of the growth of
structure using photometric LRGs \citep{cro11}. 

This paper is organized as follows.  In \S \ref{sec:method} we discuss the
method and in \S \ref{sec:data},\ref{sec:photo},\ref{sec:select} we describe
the data and sample selection. In \S \ref{sec:train} we discuss the training
set and in \S \ref{sec:results},\ref{sec:errors} we show our results, including
information about their public release, and
estimate errors.  In \S \ref{sec:usage}, we discuss the proper usage of these
results. As an example, we discuss the particular case of weak gravitational
lensing calculations.  %In section \S \ref{sec:get} we describe the fomat of the
%data to be released, and their permanent archival on the SDSS website.
Finally, in \S \ref{sec:summary} we summarize our results.

\section{Method} \label{sec:method}

The algorithm is detailed in \citet{LimaPhotoz08} and \citet{CunhaPhotoz09}.
The method is to derive weights for a training set of spectroscopically
confirmed galaxies such that the distribution of relevant quantities, such as
magnitudes or colors, matches that of a set of galaxies without known
redshifts, henceforth the photometric sample.  Assuming these quantities
correlate with redshift, and are the only relevant quantities for redshift
determination, the resulting weighted redshift histogram is proportional to the
redshift probability distribution \nofz\ of the photometric sample. 

The weighting provides a key advantage over other training sample methods such
as neural nets.  Forcing the distributions of observables of the two samples to
be proportional essentially creates a ``fair sample'' from the training set;
this approach helps avoid the biases that can arise when the training and
photometric samples have different properties.  However, this technique does
require that all areas of observable space populated by the photometric sample
are also populated by the training set, at least at some low level.

\subsection{Nearest-neighbor \pofz\ redshift estimators}\label{sec:nnpz}

\subsubsection{Weights}

In this section, we briefly review the weighting method\footnote{The weights
and \pofz\ codes are available at {\url
http://kobayashi.physics.lsa.umich.edu/$\sim$ccunha/nearest/}. Alternatively,
the code can be accessed as the git repository probwts in \url{http://github.com}} of
\cite{LimaPhotoz08}, which is required for computing \pofz.  We define the
weight, $w$, of a galaxy in the spectroscopic training set as the normalized
ratio of the density of galaxies in the photometric sample to the density of
training-set galaxies around the given galaxy.  These densities are calculated
in a local neighborhood in the space of photometric observables, e.g.,
multi-band magnitudes.  In this case, the SDSS {\it ugriz} magnitudes are our
observables; in practice we use four colors and the \rmag--band magnitude.  The
hypervolume used to estimate the density is set to be the Euclidean distance of
the galaxy to its $100^{th}$ nearest-neighbor in the training set.

The weights can be used to estimate the redshift distribution $\nwei$ of the
photometric sample:
\begin{equation}  
\nwei = \sum_{\beta=1}^{N_{\rm T}} w_\beta \delta(z-z_\beta)~.
\label{eqn:Nzest}
\end{equation}
\begin{comment}
\begin{equation}  
\nwei = \sum_{\beta=1}^{N_{\rm T,tot}} w_\beta N(z_1<z_\beta<z_2)_{\rm T},
\label{eqn:Nzest}
\end{equation}
\end{comment}

\noindent For a bin $z_1 < z < z_2$, we sum the weights of all training set
galaxies that fall within that bin.  \cite{LimaPhotoz08} and
\cite{CunhaPhotoz09} show that this indeed provides a nearly unbiased estimate
of the redshift distribution of the photometric sample, $N(z)_{\rm P}$,
provided the differences in the selection of the training and photometric
samples are solely in the observable quantities used to calculate the weights.
For example, if the photometric sample has a morphology dependent cut,
the same cut should be applied to the training sample or morphology should be one of
the observables used to measure weights.

\subsubsection{\pofz}

To estimate the redshift error distribution for each galaxy, \pofz,
we adopt the method of \cite{CunhaPhotoz09}. The \pofz\ for a given object in the
photometric sample is simply the redshift distribution of the $N$
nearest neighbors in the {\bf training} set.

\begin{equation}
\hat{P}(z) = \sum_{\beta=1}^{N_{\rm nei}} w_\beta \delta(z-z_\beta)~.
\label{eqn:pzest}
\end{equation}

\noindent This expression is the same as Eqn. \ref{eqn:Nzest} but is limited to
the nearest neighbors of a given object.  We choose $N=100$ for this study, and
estimate \pofz\ in 35 redshift bins between $z=0$ and 1.1.  We can also
construct a new estimator for $N(z)_{\rm P}$ by summing the $\hat{P}(z)$
distributions for all galaxies in the photometric sample,
\begin{equation}
\npz = \sum_{i=1}^{N_{\rm P,tot}}\hat{P}_i(z)~.
\label{eqn:Nhat2}
\end{equation}
\noindent This estimator becomes identical to that of Eqn. (\ref{eqn:Nzest})
in the limit of very large training sets.  For training sets smaller than tens
of thousands of galaxies, one can improve the \pofz s by multiplying each \pofz\ by the
ratio of $\nwei$ to $\npz$.
That is,
\begin{equation} \label{eq:pzcorrect}
P(z) \rightarrow P(z)\frac{\nwei}{\npz} \label{eqn:pzcorrect}
\end{equation}
This correction essentially corresponds to using the weights estimate as a
prior on the \pofz s.

\section{Photometric Data} \label{sec:data}

The photometric data were drawn from data release 8 (DR8) of the Sloan Digital
Sky Survey III.  Full details are given in the data release paper \citet{dr8}.
As compared to the earlier DR7 release \citep{dr7}, DR8 includes an additional
2500 deg$^2$ of new imaging in the Southern Galactic Cap (SGC), acquired to
facilitate spectroscopic target selection for the \bossfull\ (\boss), which is
part of SDSS III.

SDSS \citep{York00} images are gathered using the 2.5 meter at Apache Point
\citep{Gunn06} with the camera \citep{Gunn98} running in
time-delay-and-integrate mode.  Observations are taken in each of the SDSS
bandpasses \citep[{\it ugriz};][]{Fukugita96} nearly simultaneously as sky
moves across bands in the order $riuzg$.  The data were taken during
photometric nights under relatively good seeing conditions \citep{Hogg01}.  A
series of pipelines are run to calibrate the data
\citep{Nikhil08,Smith02,Tucker06}, derive astrometry \citep{Pier03}, and
calculate fluxes, shapes and other interesting quantities
\citep{LuptonADASS01}.  Note the calibrations used for these data are derived
using the ``ubercalibration'' technique presented in \citet{Nikhil08}.
\section{Photometric Quantities} \label{sec:photo}

In this section we describe the photometric quantities used in the creation of the
input catalog.  Most of these quantities are measured by the SDSS photometric
pipeline \photo. An early version of the pipeline is described in
\citet{LuptonADASS01};  other details can be found in the SDSS Data Release
papers, e.g. \citet{dr4} and at the SDSS III website\footnote{\sdssweb}.  We
 give a few additional details below.  In comparison to DR7, the DR8
makes use of an updated version of the \photo\ software reduction pipeline,
v5\_6 rather than v5\_4, including some updates to sky subtraction that can
change galaxy photometry and, potentially, the \pofz s.

For colors we use the SDSS ``model magnitudes'', which we refer to as
\modelmag \footnote{\DRatemags}.  Each object is fit to an elliptical
exponential disk and an elliptical \devauc\ profile convolved with a double
Gaussian approximation to the PSF model interpolated to the location of the
object \citep{LuptonADASS01,Sheldon04}.  For the \modelmag, the best fit model
in the \rmag\ band is then used to extract the flux in the other four
bandpasses, accounting appropriately for the PSF in each band. Thus the
effective aperture is the same for all bands, which is appropriate for
extraction of color information.

We use ``composite model magnitudes'' as an approximate total magnitude for
each object, which we refer to as \cmodelmag.  For each bandpass
separately, \photo\ does an additional joint fit to a non-negative
linear combination of the  best-fitting exponential and \devauc\
models. This fit determines an additional parameter {\sc frac\_deV}
($f_{deV}$), which is the fraction of the total flux estimated to come
from a \devauc\ profile.  The composite model flux in each band is then 
\begin{equation}
\textrm{Flux}_{cmodel} \equiv (1-f_{dev})\times \textrm{Flux}_{exp} + f_{dev} \times \textrm{Flux}_{dev},
\end{equation}
Because this procedure is carried out separately per band, the effective aperture for each band is
different, so these magnitudes are not appropriate for estimating colors.

For quality assurance, we use bits from the \texttt{OBJECT} bitmask output by
\photo \footnote{\DRateflags}.    We also use the \texttt{RESOLVE\_STATUS} to
choose primary observations\footnote{\DRateresolve}.  We will describe how the
flags are used in section \S \ref{sec:select}.

\begin{comment}
\begin{itemize}
  \item \texttt{SATUR}  The object contains saturated pixels.
  \item \texttt{BRIGHT} The object is very bright and must be remeasured.
  \item \texttt{DEBLEND\_TOO\_MANY\_PEAKS}
\end{itemize}
\end{comment}

\section{Photometric Sample Selection} \label{sec:select}

\subsection{Star Galaxy Separation} \label{sec:sg}

The \photo\ pipeline uses the concentration $c$ to separate stars from
galaxies.  The concentration is the difference between magnitude determined
from the best fitting PSF model \psfmag\ and the \modelmag\, which is
the better fitting of the exponential and \devauc\ models convolved
with the local PSF:
\begin{equation}
c \equiv \textrm{psfmag} - \textrm{modelmag}~.
\end{equation}
For stellar objects, the scale of the \modelmag\ approaches a delta function and
the result becomes equivalent to the \psfmag.  Thus the concentration should be
$\ge 0$ within the noise, with stars close to zero and galaxies greater than
zero.  The pipeline defines galaxies as objects with $c > 0.145$ where $c$ is
derived from the summed fluxes from all bandpasses\footnote{\DRateclass}.  

At our magnitude limit $r = $\rmax, the stellar contamination is relatively
large.  Using a small, space-based, high angular resolution data set matched to
SDSS data as a truth table, the approximate stellar contamination can be
determined.  At \rmag\ = 21 the contamination is a few percent, but the
contamination increases to approximately \contamworst\ at \rmag\ =
22\footnote{\DRsevsg}.  

For studies where completeness and purity must be known precisely,
\citet{ScrantonMag05} recommend using probabilistic star galaxy separation at
fainter mags (\rmag\ $ > $ 21); i.e.  attempt to determine the {\it
probability} that an object is a galaxy and either use that as a weight or make
appropriate cuts.

In practice the end user should choose a subset of the data that suits their
needs.  We provide a catalog here that should be a superset of objects that can
be further trimmed.

\subsection{Other Cuts}

We remove objects for which the extinction-corrected \citep{Schlegel98} model
flux is not well determined in at least one of the photometric bands.  The
adopted magnitude limits are [21, 22, 22, 20.5, 20.1] for \allmag\
respectively.

In addition to the magnitude limits described above, which ensures a reasonable
detection in at least one band, we additionally demand a detection in both the
\rmag\ and \imag\ bands.  Rather than applying a magnitude cut, we instead use
the \texttt{OBJECT} processing flags \texttt{BINNED}\{1,2,4\}, which indicate
the object was detected in the original image (binned by 1), the $\times$2
binned image, or the $\times$4 binned image, respectively\citep{Stough02}.

We remove all objects that have the following \texttt{OBJECT} flags set:
\texttt{SATUR}, \texttt{BRIGHT}, \texttt{DEBLEND\_TOO\_MANY\_PEAKS},
\texttt{PEAKCENTER}, \texttt{NOTCHECKED}, \texttt{NOPROFILE} as well as objects
that are (\texttt{BLENDED} \&\& \texttt{NODEBLEND}); in other words, detected
to be blended but not successfully deblended into components. 

We only use objects marked as \texttt{SURVEY\_PRIMARY} in their
\texttt{RESOLVE\_STATUS} flags field. Different scans on the sky image the same
objects due to the small overlap regions between adjacent scans, overlaps at
the end of the scan lines where the great circles converge, and re-observed
scan lines.  This results in duplicate observations for many objects.  These
duplicates are ``resolved'' and only a single observation is assigned
\texttt{SURVEY\_PRIMARY}.  Note this primary also implies that, if the object
is blended, it is either a child or not deblended further.  This cut is made in
the \texttt{OBJECT} flags as \texttt{!BRIGHT \&\& (!BLENDED || NODEBLEND ||
nchild == 0)}.

We require the extinction corrected \citep{Schlegel98} \cmodelmag\ in the \rmag\
band to be in the range [\rmin, \rmax].  We also restrict the extinction corrected
\modelmag\ to be within the range [\modelrmin, \modelrmax] in order to ensure
reasonable colors for the galaxies.

\begin{figure}[t] \centering
 \centering 
 \includegraphics[scale=0.6]{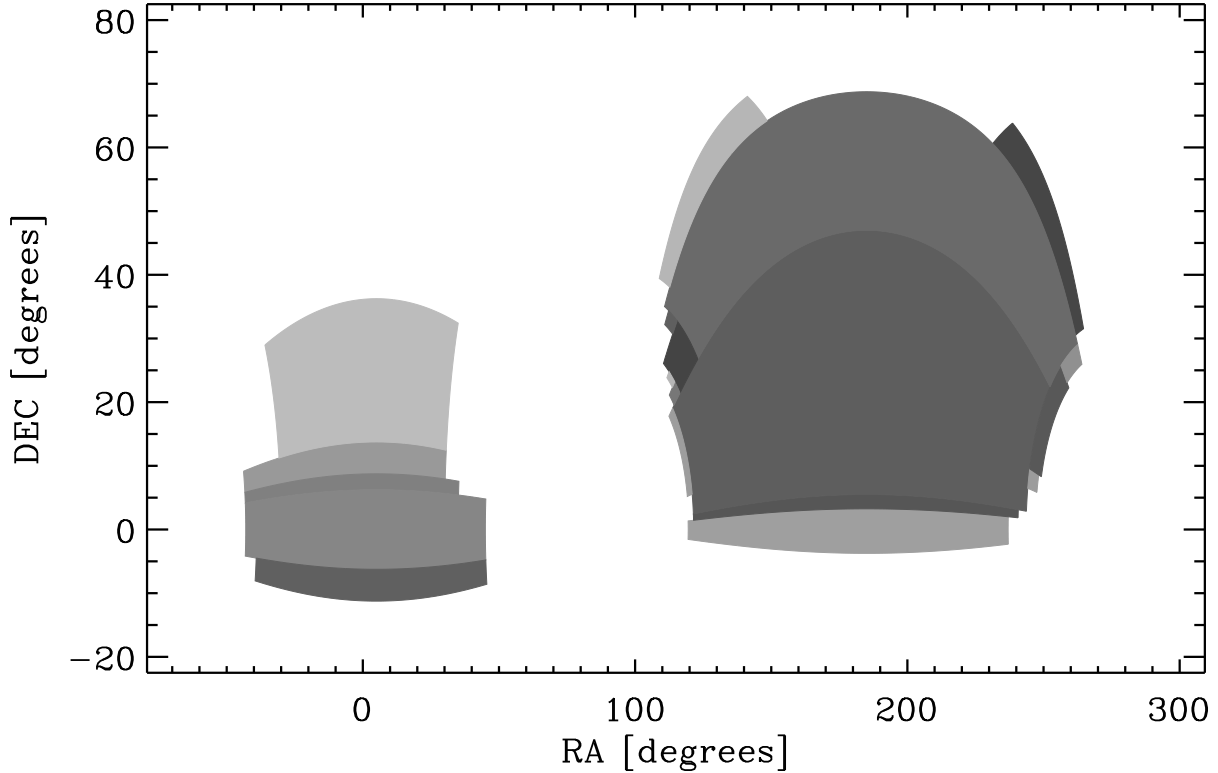}

 \caption{BOSS window function for the south galactic cap on the left and the
 north galactic cap on the right.  The differently shaded regions represent
 contiguous rectangular regions in SDSS survey coordinates, used for
 construction of the window function.  Note points with RA $>$ 300\arcdeg\ have
 been wrapped below zero to avoid the 360\arcdeg\ crossing point.}
 \label{fig:footprint}

\end{figure}

We make broad geometrical cuts on the catalog.  We trim the objects to the
\boss\ footprint, shown in Fig. \ref{fig:footprint}. We also remove any
objects near stars in the tycho2 catalog \citep{tycho2} using a variable radius
that depends on the magnitude of the star:
\begin{equation}
r = (0.0802\times B_T^2 - 1.860\times B_T + 11.625)/60.0
\end{equation}
where $B_T$ is the Tycho magnitude and $r$ is in degrees.  Finally, we remove
all objects from images taken where a \umag\ amplifier was not working\footnote{\DRsevcaveat}.

The final photometric catalog contains \nphoto\ objects.  The distributions of
extinction-corrected \rmag-band \cmodelmag\ and colors derived from
extinction-corrected \modelmag\ are shown in Fig. \ref{fig:varhist}.

\section{Training Samples} \label{sec:train}

We use a spectroscopic training set drawn from a number of sources. These
sources contain mostly galaxies and a small number of stars in order to help
characterize stellar contaminants from the photometric sample at low redshift.
In the following sections we give short details on each sample and describe our
process for matching to the photometric sample.

\subsection{Samples Used in this Study} \label{sec:train:def}

\begin{itemize} 

    \item 435,878 redshifts from the SDSS spectroscopic samples, principally
    from the \texttt{MAIN} \citep{Strauss02} and Luminous Red Galaxy
    \citep[\texttt{LRG};][]{Eisenstein01} samples, with confidence level
    \texttt{zconf}$ > 0.9$, and r-band \cmodelmag\ $ <19.5$.

    \item 445 objects from the Canadian Network for Observational Cosmology
    (CNOC) Field Galaxy Survey \cite[CNOC2;][]{yee00}\footnote{\tt
    http://www.astro.toronto.edu/$\sim$cnoc/cnoc2.html} with \texttt{Rval} $>4$
    for \texttt{Sc}$=2$ or \texttt{Rval} $> 5$ for \texttt{Sc}$=5$

    \item 151 from the Canada-France Redshift Survey
    \cite[CFRS;][]{lilly95}\footnote{\tt
    http://www.oamp.fr/people/tresse/cfrs/cfrs.html} with \texttt{Class} $\geq
    3$.

    \item 1,868 from the Deep Extragalactic Evolutionary Probe 2 survey
    \citep[DEEP2;][]{weiner05}\footnote{\tt http://deep.berkeley.edu/DR3} with
    \texttt{zqual} $\geq 3$.  Of these, 1,499 are an approximately
    magnitude-limited sample from the Extended Groth Strip (EGS).  The
    remainder is $BRI$ color-selected to target $z>0.7$ galaxies, hereafter
    denoted the non-EGS sample. 

    \item 197 from the Team Keck Redshift Survey
    \cite[TKRS;][]{wirth04}\footnote{\tt
    http://tkserver.keck.hawaii.edu/tksurvey/}.

    \item 8,633 LRGs from the 2dF-SDSS LRG and QSO Survey
    \cite[2SLAQ;][]{cannon06}\footnote{\tt http://www.2slaq.info/} with
    \texttt{qop} $\geq$ 3.

    \item  2,080 from zCOSMOS redshift survey \cite{lilly07}, with
    \texttt{cc=3.4 || 3.5 || 4.4.  || 4.5 || 9.5}.
    
    \item 1,587 from the VIMOS VLT-Deep survey
    \cite[VVDS;][]{garilli08}\footnote{\tt http://www.oamp.fr/virmos/vvds.htm}
    with \texttt{zqual} $\geq 3$.

    \item 16,874 from four fields of the PRIMUS survey
    \cite[PRIMUS;][]{coil10,cool12}\footnote{\tt
    http://cass.ucsd.edu/$\sim$acoil/primus/}.  Only PRIMUS objects with $Q$ =
    4 were used.      \end{itemize}

In table \ref{tbl:weistats} we present some statistics about each training set.

\subsection{Matching to SDSS Imaging Data} \label{sec:train:match}

We spatially match the training sets listed in \S \ref{sec:train:def} to the
photometric catalog described in \S \ref{sec:select}.  We choose the closest
match within \matchrad.  By performing this match we place the training set
galaxies on the same photometric system as the photometric set.  We also
guarantee that the matches are drawn from the same magnitude range, and have
the same quality cuts applied, as the photometric set.

As noted in \S \ref{sec:train:def}, the training sets contain some stars.
There are also stars in the photometric set, since the star galaxy separation
is not perfect.  Thus, through this matching between photometric set and
training set it should be possible to place fraction of the stars in the
photometric set at redshift zero; or at least some part of their derived \pofz.

\section{Results} \label{sec:results}

We use the algorithm described in \S \ref{sec:method} to derive weights for
each training set galaxy.  We then use these weights to calculate a weighted
redshift histogram which, under our assumptions, should be proportional to that
of the photometric set.  We also derive individual redshift probability
distributions \pofz\ for each photometric galaxy.

\subsection{Derived Weights in Observable Space}

\begin{figure}[p] \centering
    \plotone{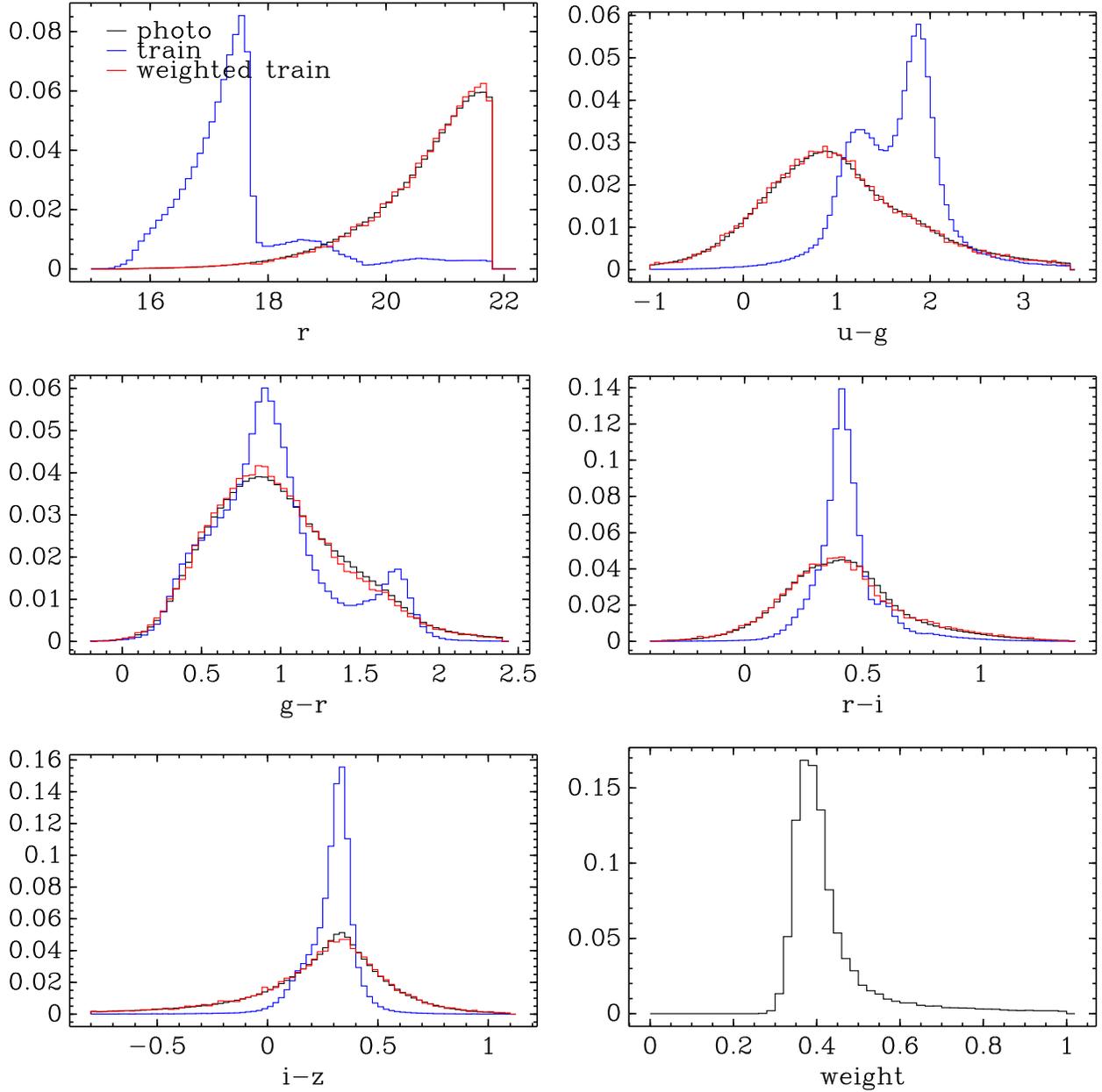}

    \caption{Distributions of photometric quantities for the photometric sample
    and training sample.  The upper left panel shows the extinction-corrected
    \rmag-band \cmodelmag.  Both samples are cut at \rmag$ < $\rmax.  Also
    shown is the weighted histogram for the training sample where the weights
    are derived to produced distributions approximately proportional to the
    photometric sample.  The following four panels show extinction-corrected
    colors based on \modelmag.  The bottom right panel shows the distribution
    of of the derived weights for the training sample. } \label{fig:varhist}

    \vspace{2em}
\end{figure}

The \rmag-band \cmodelmag\ and colors based on \modelmag\ for the photometric
and training sets are shown in Fig. \ref{fig:varhist}.  Also shown are the
derived weights for the training set and the resulting weighted histograms.
These are the fundamentally new calculations presented in this work.

The weighted training set distributions should be approximately proportional to
the photometric set distributions in order to derive good redshift
distributions.  There are deviations at \gmr\ $\sim 1.5$ and \rmi\ $\sim 0.6$,
but qualitatively the distributions are close.  We focus on the accuracy of the
recovered redshift distributions rather than a detailed comparison of these
distributions.

\subsection{Derived \nofz}

In figure \ref{fig:pofz} we present the recovered redshift distribution for the entire
\rmag\ $<$ \rmax\ sample.  Also shown is the redshift distribution of the
original training set.  These distributions are in qualitative agreement with
those shown in \citet{CunhaPhotoz09}, although that sample had a fainter
\rmag-mag limit at 22.0.  Note the sub-plot showing the region near $z=0$.  As
expected, there is a non-zero fraction of the overall distribution near redshift zero.  The
fraction of the probability at $z < 0.002$ is about 0.4\%.  It is not known exactly how many
stars are in the photometric sample, but this is probably a lower limit on the
stellar contamination (see \S \ref{sec:sg}).  We will estimate the errors on
this distribution in \S \ref{sec:errors}. These \nofz\ data are presented in
Table \ref{tab:nofz}.

\subsection{Derived \pofz}

\begin{figure}[p] \centering
    \includegraphics[scale=0.9]{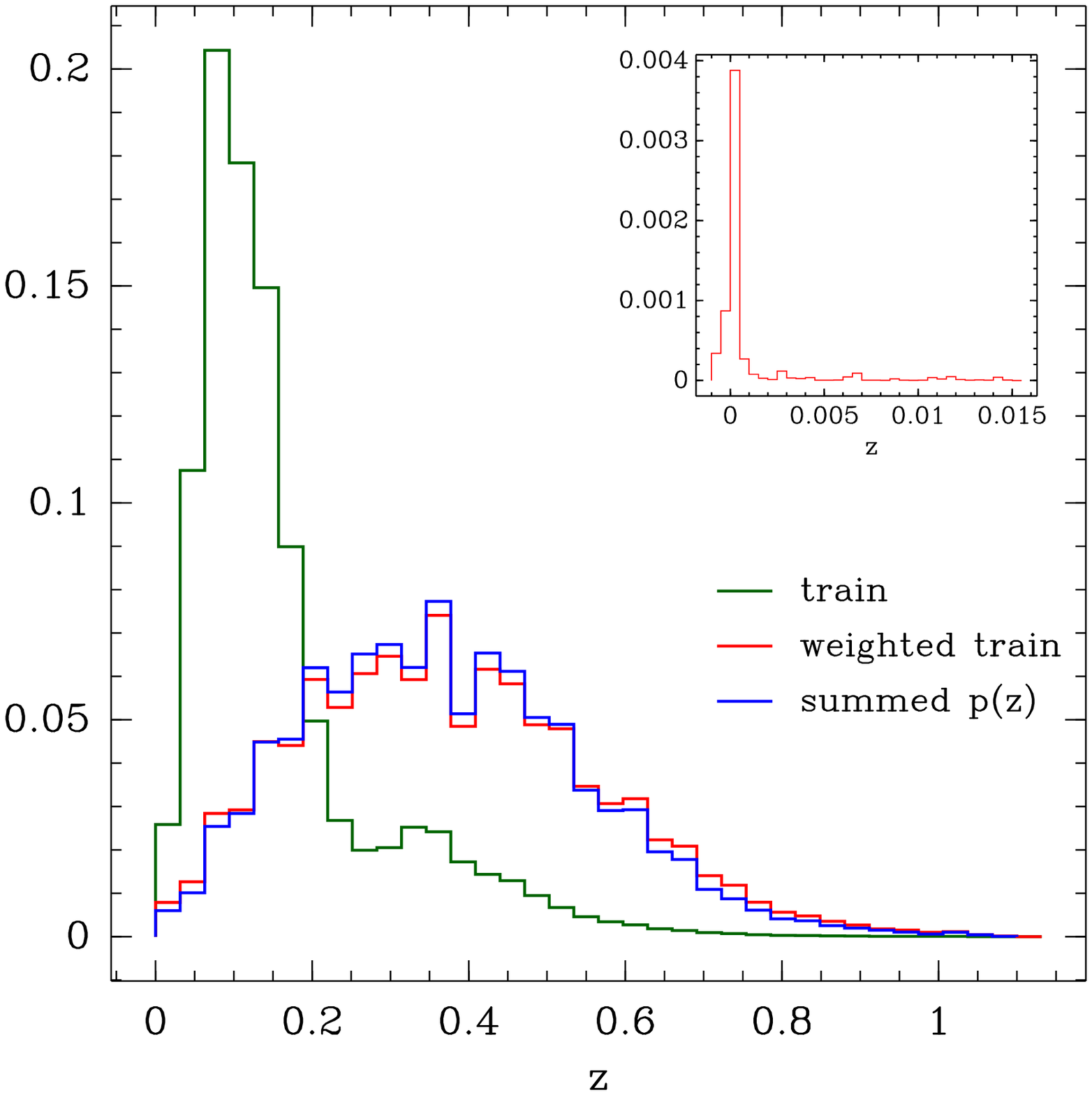}
    %\plotone{figures/zweight-10-zhist-withorig-withsum-12.eps}

    \caption{Reconstructed redshift distribution for SDSS galaxies with \rmag\
    $ < $ \rmax.  The overall reconstructed distribution, shown in red, is
    derived by creating a weighted histogram of the training set redshifts as
    described in the text.  Also shown in magenta is the sum of all \pofz\
    derived for individual galaxies.  The unweighted training set redshift
    distribution is shown in blue.  The expected errors on these distributions
    from cosmic variance in the training set is shown in Fig.
    \ref{fig:ebars}. The excess at $z \sim 0$ is due to stars in training set
    having significant weight; more detail at low redshift is shown in the
    inset.  This excess is at least partly due to the presence of real stars in
    our photometric sample resulting from imperfect star-galaxy separation.
    The fraction of the distribution at $z < 0.002$ is 0.4\%, which is probably
    a lower bound on the stellar contamination.  \label{fig:pofz}}

    \vspace{2em}
\end{figure}

Also shown in Fig. \ref{fig:pofz} is the summed \pofz\ derived for individual
galaxies.  The uncorrected $\npz$\ is, characteristically, slightly more peaked than
than $\nwei$.
In \S \ref{sec:pzcorr} we apply Eq. \ref{eqn:pzcorrect} to correct the \pofz s.

\begin{figure} [t]\centering
    \includegraphics[scale=0.7]{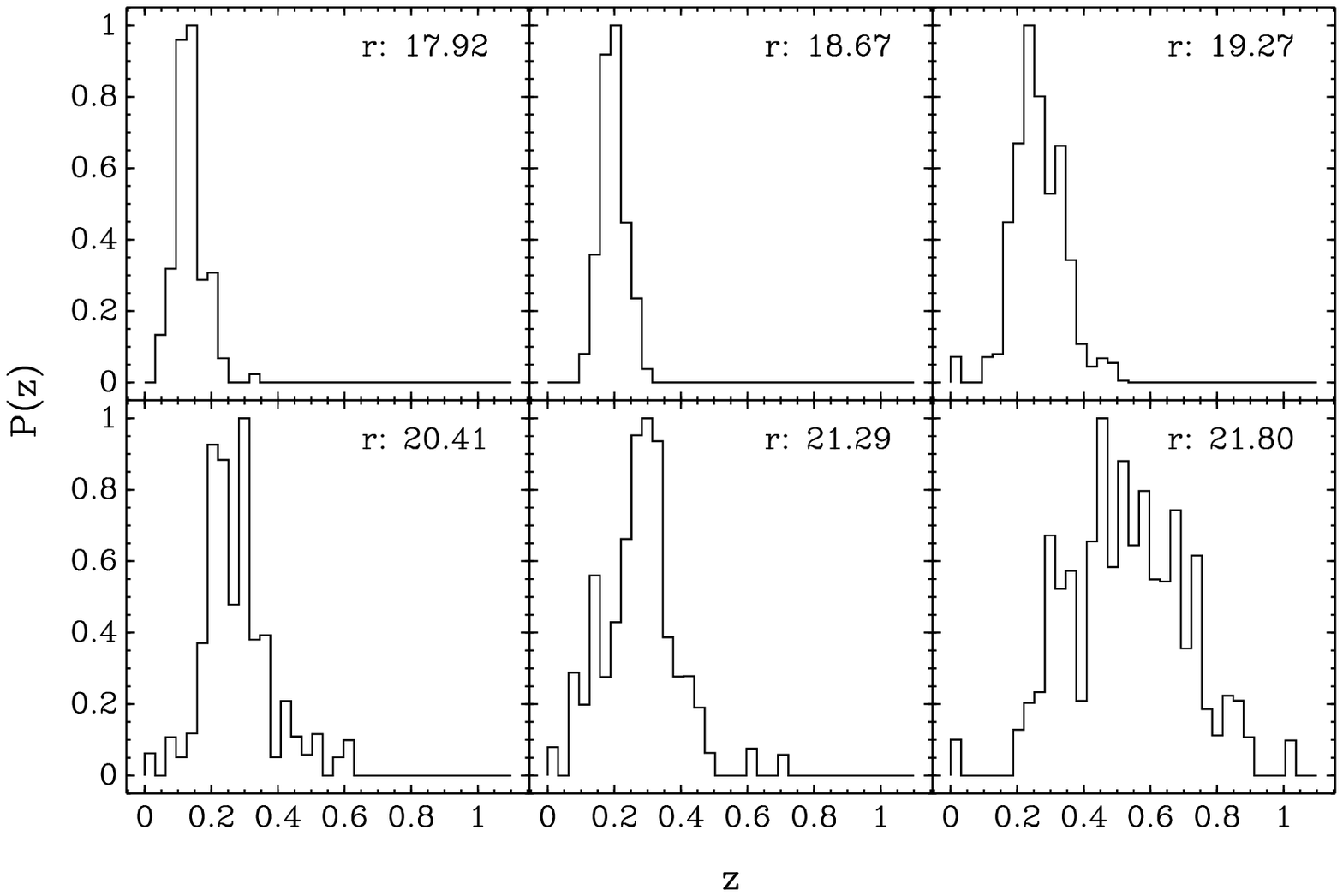}
    %\plotone{figures/11-6pofz.eps}
    \caption{Six randomly chosen \pofz s.  For each panel, an object was
    chosen from a particular magnitude range.  Column-wise from top the
    left these ranges are $r < 18$, $18 < r < 19$, $19 < r < 20$, 
    $20 < r < 21$, $21 < r < 21.5$, $21.5 < r < 21.8$.  
    The extinction-corrected $r$-band \cmodelmag\ of each object
    is indicated in the upper right of each panel.
    \label{fig:rand6pofz}}
\end{figure}

In Fig. \ref{fig:rand6pofz} we show six randomly chosen \pofz s.  Each panel
contains a \pofz\ drawn from a particular magnitude range in extinction-corrected
$r$-band \cmodelmag; these ranges are given in the figure caption.
This figure captures the general trend that the \pofz\ are broader at
fainter magnitudes, which is the expected behavior.

\begin{figure}[p]\centering
    \plotone{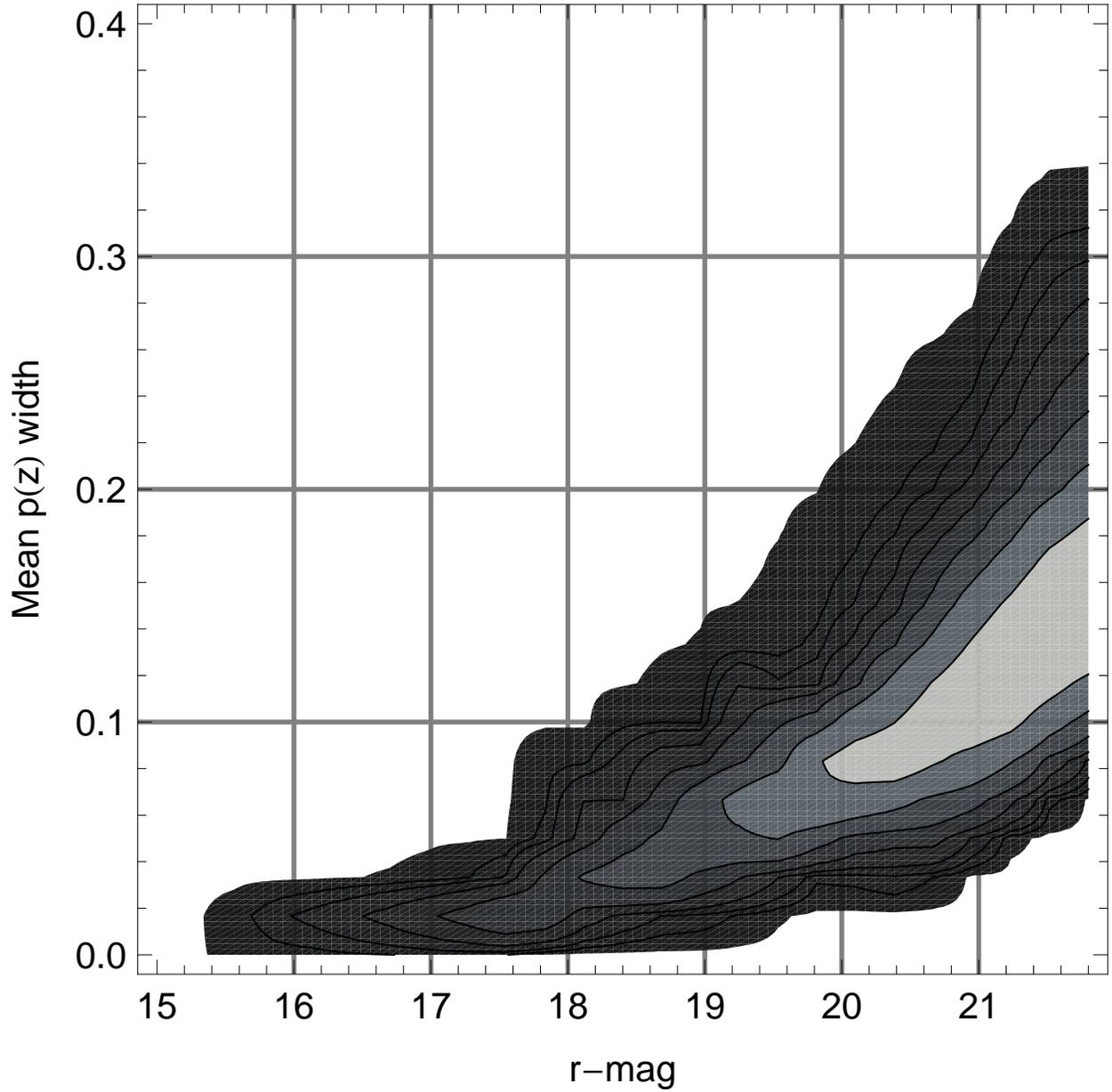}
    \caption{Density contours of the mean \pofz\ width as a function of {\it r} magnitude. 
The width of each \pofz\ is the defined as the standard deviation about the mean. 
The contours represent factors of 2 changes in density.}
    \label{fig:pzwidth}
    \vspace{2em}
\end{figure}

The uncertainty in individual \pofz s are typically dominated by shot-noise
error.  The scale of both statistical and systematic uncertainties in the
individual \pofz s is strongly correlated with the width of the \pofz\
\citep{CunhaPhotoz09}.  A broader \pofz\ reflects a larger degeneracy in
observable space, and requires more training-set objects to characterize.  Fig.
\ref{fig:pzwidth} shows the distribution of objects in the photometric sample
as a function of \rmag-band magnitude and $1 \sigma$ width of the \pofz.  The
contours indicate factor of two changes in density.  

We recommend using the $1 \sigma$ or other width measures of the \pofz\ as the
most efficient way to trim the sample for improved precision and accuracy.  The
\pofz\ width should also be a reasonable error estimator for use with other
\photoz\ methods.  However, we discourage using the peak or some other single
number statistic derived from the \pofz\ as a proxy for redshift. See 
\S \ref{sec:usage} for more details.
\begin{comment} If a single-number \photoz\ estimate is needed we recommend
using a neural network or nearest-neighbor polynomial based approach
\cite[e.g.][]{Oyaizu08}.  \end{comment}

\subsubsection{Correction to \pofz} \label{sec:pzcorr}

\begin{figure}[t]\centering
    \includegraphics[scale=0.6]{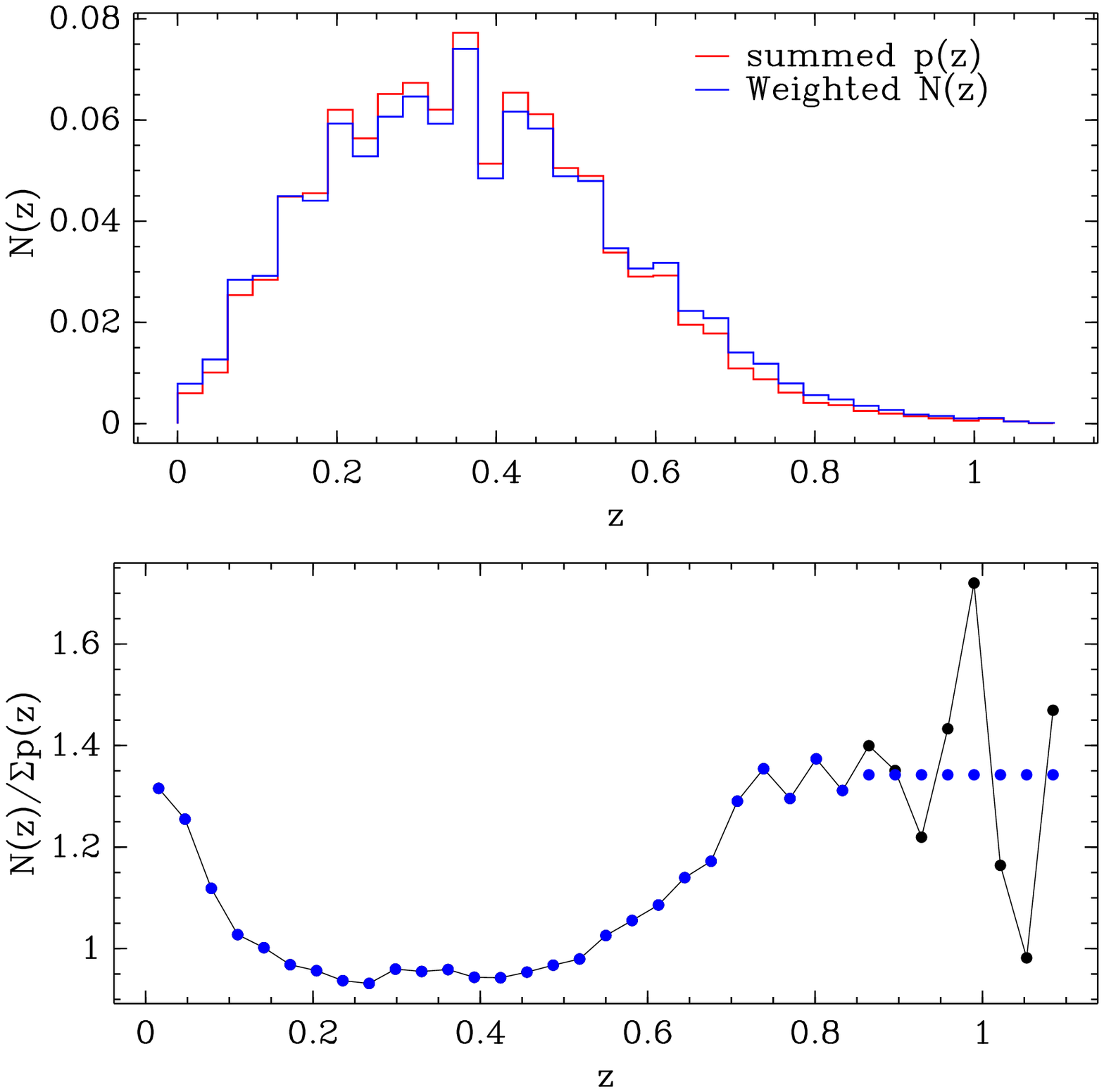}
    %\plotone{figures/pofz-correct-12.eps}

    \caption{Correction factor from Eq. \ref{eqn:pzcorrect}.  This correction
    factor is the ratio of the \nofz, which we find to be unbiased, to the summed
    \pofz\ from individual objects. The top panel shows both \nofz\ and \pofz,
    and the bottom panel is the ratio.  We apply this correction to each of the
    \pofz s in the release catalog.  Note for $z > 0.9$ we use the average
    correction from that range.}

    \label{fig:pzcorr}
    \vspace{2em}
\end{figure}

As we will demonstrate in \S \ref{sec:pofp}, the individual \pofz s are
somewhat less accurate than the overall \nofz.  We can correct the individual
\pofz\ to agree, in the mean, with the overall \nofz\ using Eqn.
\ref{eqn:pzcorrect}.  This correction factor is shown in Figure
\ref{fig:pzcorr}.  At $z \gtrsim 0.9$ neither the \nofz\ or the summed
\pofz\ are well constrained, and the correction factor is noisy.  For $z > 0.9$
we use the average correction from that range.

\subsection{Differences from previous \pofz\ derived using this method}

Unlike for the DR7 catalog, we
did not use repeat observations of our training set galaxies.  The use of
repeats can provide more localized and smoother \pofz\ estimates, and are often
useful.  However, because only part of our sample had repeat observations, the
use of repeats would effectively increase the sample variance of our results.
The use of repeats may be beneficial for LRGs because the training set is not
sample variance limited in this case.  We may release a catalog trained on
repeat observations at a future date.  

\subsection{Acquiring the Data} \label{sec:get}

The \pofz\ for all galaxies are stored on the SDSS III
website\footnote{\downloadURL}.  The data are available in both FITS format and
ASCII.  The objects are split into different files according to their SDSS run
id, with each row in the file representing the data for a single SDSS object.
The data for each object are SDSS id, the input colors and magnitude for each
object, equatorial latitude and longitude, and the estimated \pofz.

\section{Sources of Error} \label{sec:errors}

As detailed in \cite{CunhaPhotoz09}, the derived weights, and inferred \nofz,
are susceptible to at least four kinds of training-set selection effects:
spectroscopic failures, two types of large-scale structure bias (sample variance + shot
noise in the training set), and selection in non-photometric observables.  In
addition, the fact that the weights use a non-infinitesimal volume in color-magnitude
space to re-weight the photometric set can yield a small Eddington bias to the
recovered distribution.  And, as mentioned previously, incorrect star-galaxy
separation can result in incompleteness and contamination of the sample.
Because our training set consists of many different surveys with different
characteristics, it is important to quantify the contribution of each to the
overall result.  Table \ref{tbl:weistats} lists, for each of the surveys
comprising the training set, the number of objects, the approximate area, and
the fraction the survey contributes to the weighted estimate of the overall
redshift distribution.  This fraction is calculated by summing the weights
assigned to objects in each survey and dividing by the sum of weights from the
entire training set.

From Table \ref{tbl:weistats}, we see that PRIMUS carries the most weight by a
large margin at 62\%.  Overall, the magnitude-limited surveys that reach our
selection depth of \rmax\ - PRIMUS, TKRS, CNOC2, DEEP2-EGS, CFRS, VVDS, and
zCOSMOS - represent about 81$\%$ of the total weight.  
This is desirable,
because it minimizes the risk of bias in our assessment of errors in what follows.
The Table also shows that the SDSS MAIN sample ($r<17.8$) contributes only $1.7\%$ of the weights, which
is consistent with the fraction expected from simulations for a flux-limited sample 
to $r<21.8$.
The remainder of the SDSS spectra are LRGs to $r<19.4$, which make a 
contribution to the total weight at 7.4\%.

\begin{figure}[h]\centering
    \includegraphics[scale=0.7]{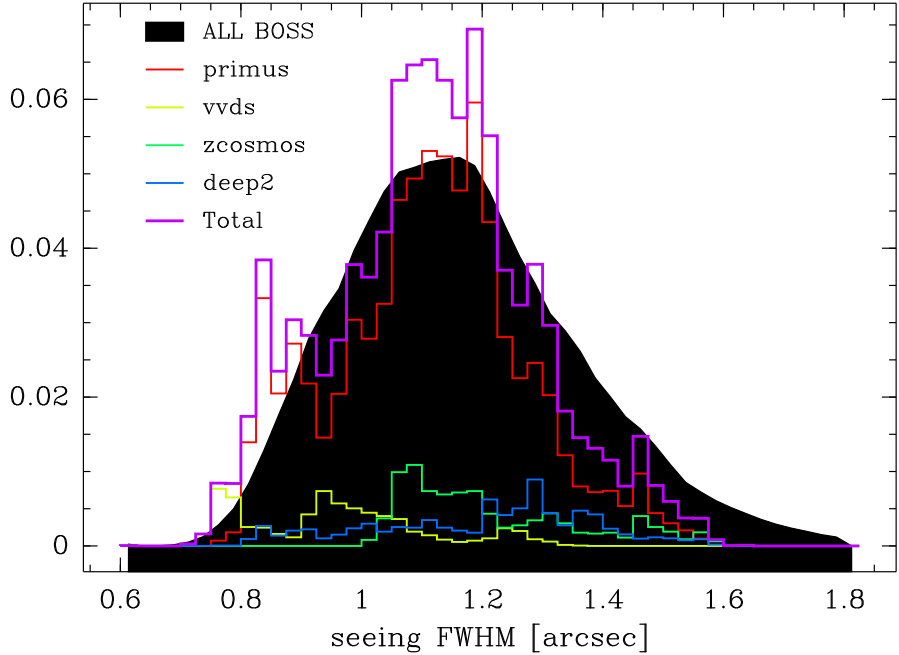}

    \caption{Distribution of seeing for the photometric sample (All BOSS) and
    the four most important training samples.  These samples are important
    because they are magnitude limited and give relatively high weight in the
    analysis.  Also shown is the sum of the training samples.  The curves for
    each sample are normalized relative to the summed curve,  and both the
    summed and photometric curves are normalized to unity.}

    \label{fig:seeing}
\end{figure}

\begin{figure}[t]\centering
    \includegraphics[scale=0.6]{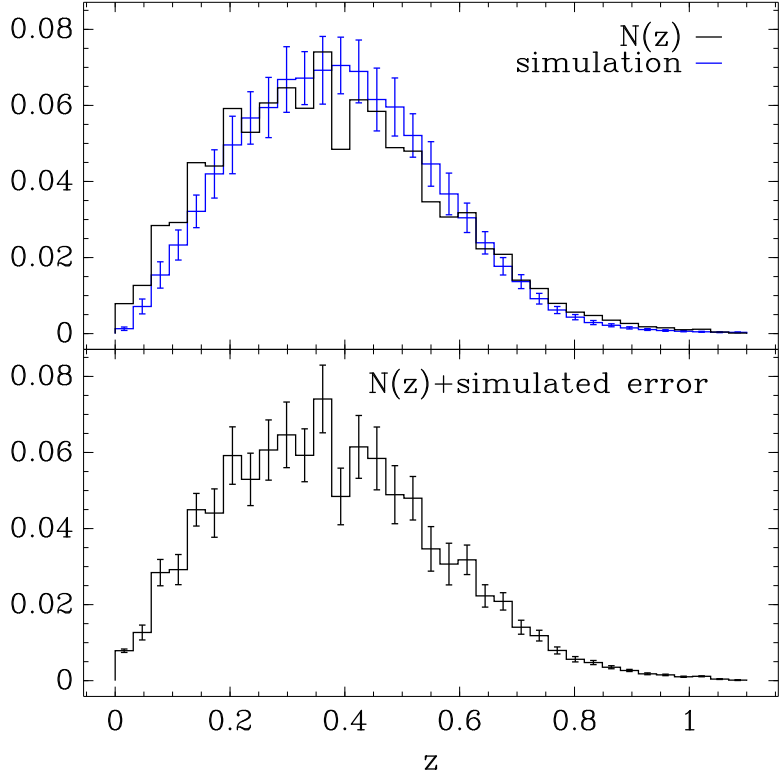}

    \caption{Top panel: Simulated redshift distribution with errors for an
    $r<$ \rmax\ sample.  The error bars are the $1 \sigma$ simulated
    variability due to sample variance in the catalogs comprising the training
    set.  Also shown is the estimated \nofz\ for our sample.  Lower panel:
    estimated \nofz\ combined with the predicted sample variance errors from
    the simulation.}

    \label{fig:ebars}
    \vspace{2em}
\end{figure}

In what follows, we identify potential sources of systematics and detail our
tests to constrain them:

\begin{itemize}

\item {\it Large-scale structure: } We expect this item to be the main source of
error.  We use galaxy+$N$-body simulations\footnote{Simulations provided courtesy of Risa Wechsler and
Michael Busha. See \cite{bushasimulations} for details.}
 to estimate the sample variance plus shot
noise uncertainties of the spectroscopic redshift distributions of the training
set.  For simplicity, we only simulate the magnitude-limited surveys of the
training set.  In addition, because of the overlap between zCOSMOS and one of
the PRIMUS fields, we neglect the zCOSMOS sample in the error estimation to
simplify the calculation.  This approach results in a $\sim$10\% increase in the error
bars relative to including zCOSMOS as an independent sample.  The predicted
error bars are overlayed on the simulated overall redshift
distribution in Fig.
\ref{fig:ebars}, and the values of the errors are given in Table \ref{tab:nofz}. The
uncertainty in the training set redshift distributions is not identical to that
of the uncertainty in the estimated redshift distributions \nofz\ derived using
the weights, hence the error bars should be thought of as approximate.
A more detailed estimation of the errors would require SDSS-specific
photometry+$N$-body simulations.  Relative to the error bars in the
training set, the error bars in the weighted \nofz\ should be (very roughly)
about 10-30\% smaller, with increased anti-correlations between neighboring
bins, but a more exact statement would require a significantly more
detailed investigation.
%There are also 
We explore these issues in more detail, and for a different data set,
in \citet{CunhaPhotozLSS11}.

\item {\it Selection in non-observables: } Two of the surveys comprising our
training set have selections in observables that are not included in the SDSS
magnitude-limited sample.  As mentioned previously, the DEEP2-nonEGS sample is
selected using BRI photometry to target galaxies above $z>0.7$.  As shown in
\citet{CunhaPhotoz09}, the use of DEEP2 in earlier versions of this catalog
resulted in a bump in the overall estimated redshift distribution around $z\sim
0.8$.  The present data release has a brighter magnitude cut and additional
training data, which has eliminated this bias.  
DEEP2-nonEGS carries about 1.4\% of the total weight.  The 2SLAQ sample targets
LRGs.  Besides SDSS magnitudes, 2SLAQ also uses morphological information in the
selection.  Because shape correlates poorly with redshift, biases due to inclusion of the
2SLAQ sample are expected to be small.  2SLAQ is an important part of our
sample because it provides a better training set for LRG's at higher redshift
than the SDSS sample.

\item {\it Spectroscopic redshift failures: } The impact of spectroscopic failures is the
most difficult to quantify.  We chose a bright $r$-magnitude cut, and relatively
stringent cuts on spectroscopic quality to minimize effects of failures, but it
is possible that, for some applications, errors due to spectroscopic failures are not negligible.
Based on the descriptions of the surveys comprising our training set, we expect the average
completeness of our training sample to be well above $90\%$.

%The characteristic of the redshift failures vary from survey to survey
%The requirements on spectroscopic success rate depend on the applications.
%The 

\item {\it Seeing: } \citet{Nakajima11} report that differences
between the seeing distribution of the galaxies in the photometric and the
training set can lead to biases in the \photoz\ error calibration.  In figure
\ref{fig:seeing} we show the seeing distributions for all of our photometric
sample compared to the four highest weight training samples, not including SDSS
for which the seeing distribution is a near perfect match.  The distributions
are qualitatively similar, but with a trend to better seeing for the training
set matches.  More quantitatively, we checked the sensitivity of our results to
seeing-induced biases by including seeing as a variable in the weights
estimation.  We find only negligible change in the recovered redshift
distribution.  Hence, although differences in seeing are in general a concern, we
find little effect in our data.
\end{itemize}

For individual \pofz s, the main source of uncertainty is shot-noise, because
only 100 galaxies were used to estimate each \pofz.  The choice to fix the
number of neighbors keeps the shot-noise equal for all galaxies, but can 
yield biases or an artificial broadening of the \pofz\ if the training set is
too sparse near the galaxy of interest.  However, we do not find the volume
spanned by the 100 nearest neighbors to be a good indicator of the \pofz\
quality, because other properties of the redshift-observable
hyper-surface affect the local density of galaxies.  A potentially more
interesting indicator of bias in individual \pofz\ s is the spatial distribution
of the training set nearest neighbors relative to the galaxy for which a \pofz\
is needed.  We leave these explorations for a future work.

%}

\section{Proper Use} \label{sec:usage}

In this section we describe the proper use of these redshift distributions.
We risk an overly pedantic discussion in order to ensure that past
mistakes in these types of analyses are not repeated.

If one desires to use the \pofz\ to evaluate any non-linear function $F(z)$,
one must integrate the function times the \pofz\ over the entire distribution;
i.e. one must take the expectation value of the function.  The reason is quite
simple. In general a function evaluated at the expectation value of $z$ does not
equal the expectation value of the function:
\begin{equation}
\langle F(z) \rangle \ne F(\langle z \rangle).
\end{equation}
The expectation value of the function should be computed as follows:
\begin{equation}
\langle F \rangle = \int_{0}^{\infty} F(z) P(z) \mathrm{d}z.
\end{equation}
It is {\bf not} correct to simply take the effective redshift $\int
z\,P(z)\,\mathrm{d}z$ and evaluate
the function at that redshift.

This statement is true in most interesting science cases.  An excellent example is in
gravitational lensing, where one must estimate the ``critical surface density''
\sigmacrit, which determines the lensing strength of a given lens-source pair; the
lensing deflection angle is proportional to \scinv.  The function
\sigmacrit\ depends on the angular diameter distances to the lens, source and
between lens and source in a non-linear manner.  The proper estimator for a lens
at redshift $z_{l}$ and source with $P(z_s)$ is
\begin{equation} \label{eq:calcscrit}
\Sigma^{-1}_{\mathrm{crit}}(z_l) = 
    \int_{0}^{\infty} \Sigma_{\mathrm{crit}}^{-1}(z_l, z_s) P(z_s) \mathrm{d}z_s.
\end{equation}

\subsection{\pofz\ and galaxy-galaxy lensing: proof-of-principle} \label{sec:pofp}

The sensitivity of observational methods to the properties of the
\pofz\ or \nofz\ depends on the details of how the observation is
carried out.  In
this section, we use the galaxy-galaxy lensing calibration method from
\cite{man08} and \citet{Nakajima11} as an example of determining this
sensitivity.  This methodology requires the use of a fair subsample of
source galaxies with spectroscopic redshifts.  For the purpose of this
paper, we use the DEEP2 EGS region, in which there are 730 galaxies
that (a) pass all cuts to be included in the SDSS source catalog from
\citet{MandelbaumSystematics05}, (b) have secure redshifts from DEEP2,
and (c) pass the additional cut $r<21.5$. 
DEEP2 EGS is only one of the many training samples used in our
analysis, 
so this exercise should be thought of as a proof-of-principle.

In brief, the quantities that we have measured are the expected calibration
bias $b_z$ in the galaxy-galaxy lensing signal due to the method of estimating
the source redshift (i.e., a multiplicative systematic error), and the degree to
which the variance in the lensing signal deviates from the ideal variance we
would achieve with optimal weighting by the true source redshift (large
deviation results in increased statistical error).  The increase in statistical
error when we have degraded redshift information arises both from source
misidentification, and also from deviations of the weights from the
optimal\footnote{Optimal weighting would also include a factor that downweights
galaxies with noisier shape measurements, $\propto (e_\mathrm{rms}^2 +
\sigma_e^2)^{-1}$.  For simplicity, we neglect this factor in the tests that
follow; however, in order to use this weighting, which modifies the effective
$N(z)$, the shape measurement error weighting must also be used in the
derivation of the \pofz\ from the training sample.} $1/\Sigma_\mathrm{crit}^2$.
Schematically, these two quantities can be determined via weighted sums over
lens-source pairs $j$ (with weight $\tilde{w}_j$; in what follows, estimated
quantities using approximate redshift information have a tilde, and ones that
use the true redshift do not):

\noindent
\begin{equation} \label{eq:lensbias}
b_z + 1 = \frac{\sum_j \tilde{w}_j (\tilde{\Sigma}_{\mathrm{crit},j}
   / \Sigma_{\mathrm{crit},j})}{\sum_j \tilde{w}_j}
\end{equation}
and
\begin{equation} \label{eq:lensweight}
\textrm{Variance ratio}  \equiv \frac{\textrm{Ideal variance}}{\textrm{Real
     variance}} = \frac{(\sum_j \sqrt{\tilde{w}_j w_j})^2}{(\sum_j
     w_j)(\sum_j \tilde{w}_j)}.
\end{equation}
For more detail, see the aforementioned papers.

\begin{figure} [h]\centering
    \includegraphics[scale=0.5]{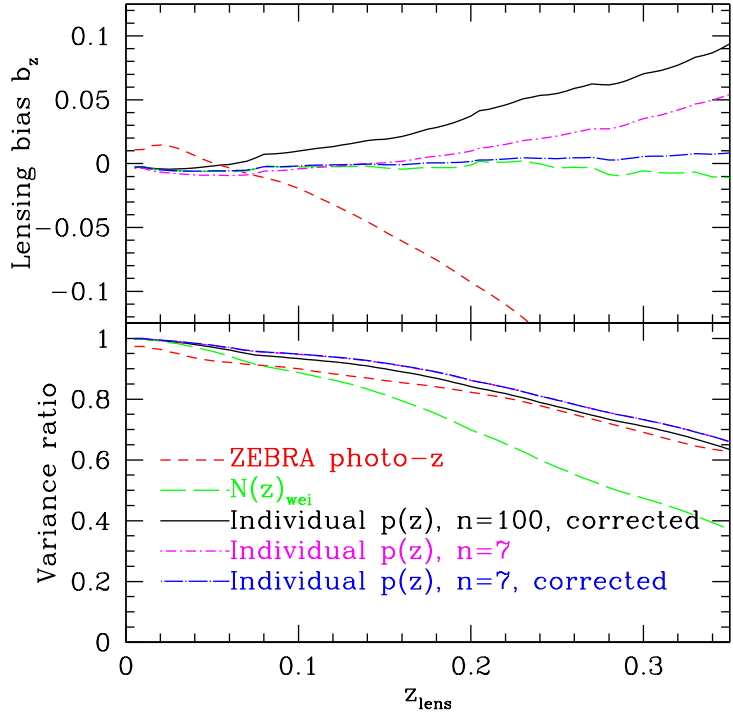}

    \caption{ Proof-of-concept analysis of errors in a fictitious lensing
    analysis.  For this example we used only DEEP2-EGS galaxies but perfect
    weights estimate; the sample variance and width of individual \pofz s are
    much larger than for the DR8 analysis.  {\em Top:} Lensing signal
    calibration bias (Eq.  \ref{eq:lensbias}) as a function of lens redshift,
    for four  cases labeled on the plot and discussed in the text.  {\em
    Bottom:}  Ratio of the ideal to the real signal variance when using
    different  methods of redshift determination; the goal is to stay as close
    to unity as possible. \label{fig:simplebias}}

\end{figure}

In Figure \ref{fig:simplebias} we show the results of these calculations for
several test cases.  First, the red short-dashed curve provides, as a baseline,
the calibration bias (top) and variance ratio (bottom) when using the ZEBRA
\photoz\ studied in \citet{Nakajima11}.  As shown, there is a
significant bias in the lensing signal that must be calibrated.
Next, the green long-dashed line shows what happens if we use the
$N(z)_\mathrm{wei}$ as an estimate of the redshift distribution, rather than
using any individual galaxy \photoz\ or \pofz\ information.  Crucially, the
lensing signal is unbiased in this case. However, as shown in the bottom
panel, we do find an increased
statistical error due to lack of redshift information on a per-galaxy basis.

Third, the solid black line demonstrates what happens when we use the individual
\pofz s to estimate $\Sigma_\mathrm{crit}$ using Eq. \ref{eq:calcscrit}.  These
\pofz s are derived from a very specific, idealized case, using {\em only} EGS
both as the training sample and the photometric sample.  In this case, the
individual \pofz s are on average $40\%$ broader than the DR8 \pofz s because of
the use of 100 neighbors to construct each \pofz\ when the training sample
itself is only $7$ times as large.  To compensate for the bias introduced by
the small size of the training sample, we have imposed a multiplicative
correction factor to the \pofz s such that $\sum P(z) = N(z)_\mathrm{wei}$ using
Eq. \ref{eq:pzcorrect}.  Nonetheless, there is a calibration bias due to the
very significant width of the \pofz s (which can be removed using a calibration
sample); but the variance ratio is still far closer to optimal than when we did
not use weighting information, and slightly closer than when we used ZEBRA
\photoz.  

The magenta dot-dashed line shows the results when 7 neighbors are used to
estimate the \pofz, not including the galaxy itself. The blue dot-long dashed
line shows the same case but with the Eq.~\ref{eq:pzcorrect} correction.  This
use of 7 neighbors reduces the abnormally broad \pofz s caused by using such a
small training sample and 100 neighbors.  The mean \pofz\ width for the 7
neighbors case is 0.0989, to be compared to the mean width of the DR8 \pofz s of
0.0983. The calibration bias for 7 neighbors is also quite close to the ideal
case with $N(z)_\mathrm{wei}$, and the weighting is the closest to optimal of
all the cases considered in this paper.

To summarize, we have demonstrated for this simplified training set that, for
the purpose of lensing, we achieve a perfect signal calibration when using
$N(z)_\mathrm{wei}$; i.e., no individual galaxy redshift information.  However,
the weighting is suboptimal.  When we use individual \pofz s, the lensing signal
can be biased due to their finite width even if $\sum P(z) =
N(z)_\mathrm{wei}$, but this bias can be calibrated.  The advantage of
using individual \pofz\ information is that statistical errors on the lensing
signal are reduced due to more optimal weighting.  This is because a
signal-to-noise ratio weighting is proportional to $\langle \Sigma_{\mathrm{crit}}^{-2}
\rangle$, so sources expected to be behind the lens are given higher weight
than those expected to be close to or in front of the lens.

Again, we emphasize that this analysis used only DEEP2 EGS, and should be used
as a proof-of-principle to gain intuition.  Users of these data should perform
similar analyses to these but matched to their exact analysis and selection
criteria.

\section{Summary} \label{sec:summary}

In this paper we presented a catalog of photometric redshift probability
distributions for the SDSS DR8.  With some modifications, our method is the
same as that used to generate the \pofz\ catalog for SDSS DR7, presented in
\cite{CunhaPhotoz09}.  For this catalog, we used the ubercal photometry
\citep{Nikhil08}.  We also included the PRIMUS galaxy sample, which more than
doubles the number of galaxies in our training set that are drawn from a
flux-limited sample other than SDSS.  The addition of PRIMUS provided a
significant increase in the total area of the non-SDSS training set, which
reduces the sample variance.  We examined several potential sources of error,
including shot noise, sample variance, seeing, star-galaxy separation, and
spectroscopic failures.  We expect that sample variance is the main source of
uncertainty in our overall redshift distribution.  For individual \pofz s,
shot-noise is the limiting uncertainty, since each \pofz\ is based on 100
training set galaxies.  These \pofz s, and the ensemble \nofz\ derived in this
work (Table~\ref{tab:nofz}), should be useful for a variety of science
applications, such as galaxy angular two-point correlation functions, galaxy
cluster detection and weak gravitational lensing.

\section*{Acknowledgments}

ES is supported by DOE grant DE-AC02-98CH10886.  CC was supported by DOE OJI
grant under contract DEFG02-95ER40899 and the Kavli Fellowship at Stanford.

Thanks to Don Schneider for a careful reading of the manuscript and many
helpful suggestions.

Funding for the DEEP2 survey has been provided by NSF grants AST95-09298,
AST-0071048, AST-0071198, AST-0507428, and AST-0507483 as well as NASA LTSA
grant NNG04GC89G. 

A portion of the data presented herein were obtained at the W. M. Keck
Observatory, which is operated as a scientific partnership among the California
Institute of Technology, the University of California and the National
Aeronautics and Space Administration. The Observatory was made possible by the
generous financial support of the W. M. Keck Foundation. The DEEP2 team and
Keck Observatory acknowledge the very significant cultural role and reverence
that the summit of Mauna Kea has always had within the indigenous Hawaiian
community and appreciate the opportunity to conduct observations from this
mountain.  Funding for the SDSS and SDSS-II has been provided by the Alfred P.
Sloan Foundation, the Participating Institutions, the National Science
Foundation, the U.S. Department of Energy, the National Aeronautics and Space
Administration, the Japanese Monbukagakusho, the Max Planck Society, and the
Higher Education Funding Council for England. The SDSS Web Site is
http://www.sdss.org/.

The SDSS is managed by the Astrophysical Research Consortium for the
Participating Institutions. The Participating Institutions are the American
Museum of Natural History, Astrophysical Institute Potsdam, University of
Basel, University of Cambridge, Case Western Reserve University, University of
Chicago, Drexel University, Fermilab, the Institute for Advanced Study, the
Japan Participation Group, Johns Hopkins University, the Joint Institute for
Nuclear Astrophysics, the Kavli Institute for Particle Astrophysics and
Cosmology, the Korean Scientist Group, the Chinese Academy of Sciences
(LAMOST), Los Alamos National Laboratory, the Max-Planck-Institute for
Astronomy (MPIA), the Max-Planck-Institute for Astrophysics (MPA), New Mexico
State University, Ohio State University, University of Pittsburgh, University
of Portsmouth, Princeton University, the United States Naval Observatory, and
the University of Washington.

%Summarize what we did.  Compare to other estimators?

%- Compare to DR7 p(z)'s: Added PRIMUS, no repeats-> focus on decreasing sample variance, UBERCAL, Payed attention to the seeing.

%- Previous usage in the literature. Mandelbaum et al, Carnero et al, Crocce et al (, Liu et al, 

%- Abrahamse: mathematical proof that P(z)'s are unbiased if N(z) is correct.

\bibliographystyle{apj}
% Bib database
\bibliography{apj-jour,astroref}

\begin{deluxetable}{lrrr}
\tablecaption{Statistics for Each Training Set \label{tbl:weistats}}
\tablewidth{0pt}
\tablehead{
    \colhead{Survey} &
    \colhead{Number} &
    \colhead{Area} &
    \colhead{Weight Fraction} \\
    &
    of Objects &
    (sq. deg.) &
}
\
\startdata
PRIMUS$^*$              & 16,874   & 5.2     & 0.63      \\
zCOSMOS$^*$             & 2,080    & 1.7     & 0.075     \\
SDSS DR5                & 435,875  & 5740    & 0.074     \\
2SLAQ                   & 8,633    & 180     & 0.060     \\
VVDS$^*$                & 1,587    & 4.0     & 0.060     \\
DEEP2-EGS$^{*}$         & 1,499    & 0.4     & 0.058     \\
SDSS DR5 ($r < 17.8$)   & 376,625  & 5740    & 0.017     \\
CNOC2$^*$               & 445      & 0.4     & 0.016     \\
DEEP2-nonEGS$^{*}$      & 369      & 2.8     & 0.014     \\
CFRS$^*$                & 151      & $<$0.1  & 0.0076    \\
TKRS$^*$                & 197      & 0.07    & 0.0055
\enddata
\tablecomments{Number of galaxies, area in square degrees, and fractional contribution to the 
weights estimate of $N(z)$. The ``*'' indicates samples that are approximately flux-limited to our
selection depth. }
\end{deluxetable}

\begin{deluxetable}{cccc}
\tablecaption{Estimated \nofz\ and Sample Variance Errors \label{tab:nofz}}
\tablewidth{0pt}
\tablehead{
    \colhead{$z_{min}$} &
    \colhead{$z_{max}$} &
    \colhead{\nofz} &
    \colhead{Sample Variance} \\
    &
    &
    &
    Error
}
\
\startdata
0.000 & 0.031 & 0.788 & 0.045 \\
0.031 & 0.063 & 1.267 & 0.195 \\
0.063 & 0.094 & 2.841 & 0.346 \\
0.094 & 0.126 & 2.921 & 0.396 \\
0.126 & 0.157 & 4.496 & 0.429 \\
0.157 & 0.189 & 4.407 & 0.636 \\
0.189 & 0.220 & 5.920 & 0.756 \\
0.220 & 0.251 & 5.293 & 0.689 \\
0.251 & 0.283 & 6.065 & 0.791 \\
0.283 & 0.314 & 6.464 & 0.863 \\
0.314 & 0.346 & 5.926 & 0.696 \\
0.346 & 0.377 & 7.407 & 0.889 \\
0.377 & 0.409 & 4.845 & 0.745 \\
0.409 & 0.440 & 6.147 & 0.827 \\
0.440 & 0.471 & 5.845 & 0.827 \\
0.471 & 0.503 & 4.890 & 0.763 \\
0.503 & 0.534 & 4.797 & 0.572 \\
0.534 & 0.566 & 3.466 & 0.586 \\
0.566 & 0.597 & 3.066 & 0.548 \\
0.597 & 0.629 & 3.178 & 0.388 \\
0.629 & 0.660 & 2.229 & 0.293 \\
0.660 & 0.691 & 2.085 & 0.228 \\
0.691 & 0.723 & 1.406 & 0.183 \\
0.723 & 0.754 & 1.185 & 0.141 \\
0.754 & 0.786 & 0.795 & 0.093 \\
0.786 & 0.817 & 0.564 & 0.070 \\
0.817 & 0.849 & 0.477 & 0.054 \\
0.849 & 0.880 & 0.354 & 0.039 \\
0.880 & 0.911 & 0.268 & 0.029 \\
0.911 & 0.943 & 0.181 & 0.024 \\
0.943 & 0.974 & 0.152 & 0.020 \\
0.974 & 1.006 & 0.103 & 0.016 \\
1.006 & 1.037 & 0.115 & 0.013 \\
1.037 & 1.069 & 0.042 & 0.011 \\
1.069 & 1.100 & 0.015 & 0.010
\enddata

\tablecomments{Reconstructed redshift distribution \nofz\ for SDSS galaxies
with \rmag\ $ < $ \rmax.  The first two columns specify the redshift range of
the bin and the third is the reconstructed \nofz, with arbitrary normalization.
The fourth is the sample variance errors on \nofz\ derived from simulations,
which we expect to be the dominant uncertainty.  These sample variance errors
should be thought of as a rough estimate.  A more perfect match would require a
simulation more specifically tuned to the SDSS data.}

\end{deluxetable}

\end{document}

%% file: newcommands.tex
\newcommand{\boss}{\texttt{BOSS}}
\newcommand{\bossfull}{Baryon Oscillation Spectroscopic Survey}

\newcommand{\sdssweb}{http://www.sdss3.org}

\newcommand{\DRatemags}{http://www.sdss3.org/dr8/algorithms/magnitudes.php}
\newcommand{\DRateflags}{http://www.sdss3.org/dr8/algorithms/flags\_detail.php}
\newcommand{\DRateresolve}{http://www.sdss3.org/dr8/algorithms/resolve.php}
\newcommand{\DRateclass}{http://www.sdss3.org/dr8/algorithms/classify.php}
\newcommand{\DRsevsg}{http://www.sdss.org/DR7/products/general/stargalsep.html}
\newcommand{\DRsevcaveat}{http://www.sdss.org/dr7.1/start/aboutdr7.1.html\#imcaveat}

\newcommand{\umag}{$u$}

\newcommand{\rmag}{$r$}
\newcommand{\imag}{$i$}

\newcommand{\allmag}{$u,g,r,i,z$}

\newcommand{\gmr}{$g-r$}
\newcommand{\rmi}{$r-i$}

\newcommand{\psfmag}{\texttt{psfmag}}
\newcommand{\modelmag}{\texttt{modelmag}}
\newcommand{\cmodelmag}{\texttt{cmodelmag}}

\newcommand{\devauc}{De Vaucouleurs'}

\newcommand{\sigmacrit}{$\Sigma_{\mathrm{crit}}$}
\newcommand{\scinv}{$\Sigma^{-1}_{\mathrm{crit}}$}

\newcommand{\photoz}{photo-$z$}

\newcommand{\photo}{\texttt{PHOTO}}

%% file: authors.tex
\author{
Erin S. Sheldon,\altaffilmark{1}
Carlos Cunha,\altaffilmark{2}
Rachel Mandelbaum,\altaffilmark{3,4}
J. Brinkmann,\altaffilmark{5}
and Benjamin A. Weaver\altaffilmark{6}
}

\altaffiltext{1}{Brookhaven National Laboratory, Bldg 510, Upton, New York 11973}
\altaffiltext{2}{Department of Physics, University of Michigan, 500 East University, Ann Arbor, MI 48109-1120.}
\altaffiltext{3}{Princeton University Observatory, Peyton Hall, Princeton, NJ 08544.}
\altaffiltext{4}{Department of Physics, Carnegie Mellon University, 5000 Forbes Avenue, Pittsburgh, PA 15213.}
\altaffiltext{5}{Apache Point Observatory, P.O. Box 59, Sunspot, NM 88349.}
\altaffiltext{6}{Center for Cosmology and Particle Physics, Department of Physics, New York University, 4 Washington Place, New York, NY 10003.}